\documentclass[linenumbers]{aastex631}
\usepackage[mathlines]{lineno}
\usepackage{bbm}
\usepackage{mathrsfs}
\usepackage{amssymb,amsmath}
\usepackage{rotating}
\usepackage{color}

\newcommand\al{\alpha}

\newcommand\de{\delta}
\newcommand\ep{\epsilon}

\renewcommand\th{\theta}

\newcommand\rh{\rho}

\newcommand\<{\langle}
\renewcommand\>{\rangle}

\newcommand\ie{\emph{i.e.}}
\newcommand\eg{\emph{e.g.}}

\newcommand\beq{\begin{equation}}
\newcommand\eeq{\end{equation}}
\newcommand\bea{\begin{eqnarray}}
\newcommand\eea{\end{eqnarray}}
\newcommand\bal{\begin{align}}
\newcommand\eal{\end{align}}

\newcommand\X{\times}
\newcommand\fr{\frac}

\renewcommand\d{\mathrm{d}}
\newcommand\e{\mathrm{e}}
\newcommand\h{\mathbbm{h}}
\renewcommand\i{\mathrm{i}}

\newcommand\R{\mathbb{R}}
\newcommand\C{\mathbb{C}}
\newcommand\Z{\mathbb{Z}}

\renewcommand\bal{\mbox{\boldmath$\alpha$}}

\begin{document}

\title{Are dark matter and dark energy omnipresent?}
\nolinenumbers

\begin{abstract}
\nolinenumbers
A set of temporal singularities (transients) in the mass-energy
density and pressure, bearing a specific mathematical structure which represents a new solution to the continuity equation
(\ie~conservation of mass-energy) and satisfying the strong energy condition, is proposed to account for the expansion history of a homogeneous Universe, and
the formation and binding of large scale structures as a continuum approximation of their cumulative
effects. These singularities are unobservable because they occur rarely in time and are unresolvably
fast, and that could be the reason why dark matter and dark energy have not been found.   Implication on
inflationary cosmology is discussed.  The origin of these temporal singularities is unknown, safe to say that the same is true of the moment of the Big Bang itself.
This work complements a recent paper, where a topological
defect in the form of a spatial, spherical shell of density singularity giving rise to a 1/r attractive force
(to test particles of positive mass) but zero integrated mass over a large volume of space, was proposed
to solve the dark matter problem in bound structures but not cosmic expansion. The idea also involved a
negative density, which is not present in the current model.
\end{abstract}


\author{Richard Lieu}
\affiliation{Department of Physics and Astronomy, University of Alabama, Huntsville, AL 35899}
\section{Introduction}

Under the assumption of a spatially flat, homogeneous, and isotropic Universe, an appropriate line element for which is of the Friedmann-Robertson-Walker form 
\beq ds^2 = c^2 dt^2 - a^2 (t) (dr^2 + r^2 d\th^2 + r^2 \sin^2 \th d\phi^2), \label{FRW} \eeq 
the dimensionless `expansion factor' is completely and consistently determined by substituting the metric tensor for (\ref{FRW}) into the Einstein field equations.  Specifically, the ensuing pair of independent equations, known as the Friedmann equations, are of the form 
\beq \left(\fr{\dot a}{a}\right)^2 = \fr{8\pi G\rh}{3} +\fr{K}{a^2};~\fr{\ddot a}{a} = -\fr{4\pi G}{3} (\rh+3p) \label{Friedmann} \eeq 
where $\rh$ and $p$ are the density and pressure of the cosmic substrata.  For a single fluid Universe, the ratio of $w=p/\rh$ is a dimensionless constant known as the equation of state parameter.

Although the FRW metric and the Friedmann equations form the backbone of modern Cosmology, there are significant imperfections in the observational verification of its curremt solution which at least warrant the search for alternatives.  

The lowest order problems are the nature of dark matter and dark energy, namely in order to account for the temperature anisotropy of the cosmic microwave background (CMB) \citep{Efstathiou1992,Bennett2003} and the SN1a extension of the Hubble diagram \citep{Riess1998,Perlmutter1999,Brout2022} the cosmic substrata at $z=0$ comprises $69.2 \pm 1.2$~\% by mass of a component with $w\approx -1$, formerly referred to as the Cosmological constant and currently dark energy, $\approx 25.8 \pm 1.1$~\% of a component with $w\approx 0$ known as dark matter, $4.84 \pm 0.10$~\% are baryons with $w\approx 0$, and a negligible contribution from radiation; see Table 9 of \cite{pla16}.  The identities of both substrates remain unknown, with the dark matter problem being especially acute because it has been exactly 100 years since the first documented suggestion of the existence of such particles was mad \citep{Kapteyn1922}, yet they have not been found.

The next order problem is termed `Hubble tension', which, tersely speaking, is the discrepancy between the SN1a measurements at relatively low redshift $z$  and the CMB measurements at high $z$ of the Hubble constant $H_0$, with the former yielding values of $H_0$ in the range $72 -- 74$ \citep{Riess2022} and the latter $67 -- 68$ \citep{Alves2020}, both expressed in units of km~s$^{-1}$~Mpc$^{-1}$.  

The purpose of this paper is to propose another possible solution to the lowest order problems of dark matter and dark energy, by showing that there exists a class of temporal singularities capable of driving an expansion history of the Universe at least from the epoch of matter radiation equality onward, and potentially of relevance to the horizon problem and structure formation as well.  Our consideration stems from the premise that a time-inhomogeneous Universe could possess multiple singularities beyond the Big Bang and cyclical cosmology, to include the possibility of discretization in the expansion rate; more precisely, the expansion factor $a(t)$ does not increase continuously with $t$ but in steps.  It will be shown in this way that the observed expansion history of the Universe since matter-radiation equality, \ie~ during the epochs of deceleration and acceleration, can be reproduced without invoking dark matter and dark energy.

\section{Equation of state; difficulty with negative density}

In an earlier paper, a topological defect in the form of massless spherical shells was proposed to explain how large-scale structures, namely galaxies, groups, and clusters of galaxies, could remain as gravitationally bound systems even if they do not possess dark matter \citep{Lieu2024}.  At that time, the emphasis was on the binding of structures; as a result, no consideration was given to the cosmological expansion of space.  Since structures which condensed out of the Hubble flow of the cosmic substratum have observable density gradients are hence a preferred location, singularities in space were enlisted to account for the existence in lieu of dark matter.   

However, there was a controversy about the existence of negative density in the spherical shells.  Specifically the density of each of the many concentric shells postulated to exist in a galaxy, group, or cluster of galaxies was of the form \beq \rh (r)= \fr{A}{r^2} [\de (r-R)+r\de' (r-R)], \label{rshell} \eeq which leads to an integrated mass over any spherical volume of radius $r$ and centered at the origin of \beq M (r)= \int_0^r 4\pi r'^2 \rho (r') dr' = 4\pi Ar\de (r-R). \label{Mi} \eeq Hence the total enclosed mass $M$ is only non-vanishing at $r=R$, ie~the vast majority of the space is empty (apart from baryons and radiation).  Yet the acceleration is, by Poisson equation, $F=-d\Phi/dr = -4\pi GA\de (r-R)/r$ which, being inversely proportional to $r$, yields the key signature of a flat rotation velocity curve for circular stellar orbits in a galaxy for stars
on the mass shell\footnote{The term `on mass shell' is not to be confused with a similar terminology in quantum electrodynamics, which has a very different meaning.} $r=R$, with the quantity $4\pi A$ having the meaning of the equivalent underlying total mass responsible for the flat rotation.  A weakness of this model is the presence of negative density arising from the $\de ' (r-R)$ term of (\ref{rshell}) at $r=R+\ep$ when the delta function derivative is interpreted in the usual manner as the limiting case of the slope of a Gaussian profile centred at $r=R$.


There is a possible way out of the difficulty of the negative mass, however.  
It utilises the general version of the spherically symmetric Poisson equation 
\beq \nabla^2 \Phi =\fr{1}{r^2}\fr{d}{dr} \left(r^2 \fr{d\Phi}{dr}\right) = \rho + 3p, \label{GP} \eeq 
where $p$ is the pressure of the fluid, by the following re-attribution of (\ref{rshell}), 
\beq \rho (r)= \fr{A}{r^2} \de (r-R);~p(r) = \fr{A}{3r} \de' (r-R);~A>0. \label{denpre} \eeq  
In this way, the mass-energy density $\rho (r)$ (\ie~the time-time component of the stress-energy tensor) is positive definite, but the pressure is negative (which is acceptable) for $r\to R^+$.  

To calculate the pressure to density ratio of the fluid inside a large scale bound structure due to an ensemble of concentric shells one may write, for the entire system, 
\beq
    \rho_S (r)  = \fr{1}{r^2}\sum_j A_j  \de (r-R_j);~p_S (r) = \fr{1}{3r} \sum_j A_j \de' (r-R_j);~A_j >0~{\rm for~all}~j. \label{S}
\eeq 
Assuming the shell spacing between consecutive pairs is much smaller than the radius of the shell, one may replace the summation over shells by an integral, namely 
\beq \rho_S (r) = \fr{A(r)n(r)}{r^2};~p_S (r) = \fr{1}{3r} \fr{d}{dr} [(n(r) A(r)],  \label{IS} \eeq 
where $n(r) dr$ is the number of shells having radii between $r$ and $r+dr$.  In this way one sees that for slowly varying  $n(r) A(r)$ (a criterion satisfied by \citet{Lieu2024}) the pressure of the fluid $p_S (r)$ is negligible compared to the density $\rho_S$, \ie~the fluid is in this respect similar to cold dark matter.  The key difference is that (\ref{IS}) obeys a pressureless flow equation  \beq \dot\rho_S = -\fr{2\dot r}{r}
(\rho_S + p_S)  \label{2Dflow}  \eeq where the right side carries the factor $2\dot r/r$ rather than the usual $3\dot r/r$ because the shells are two-dimensional surface structures which undergo only a change in area when the radius is perturbed (shell thickness remains unresolvably small). 

Unfortunately (\ref{2Dflow}) also spells out the other shortcoming of this model. Although the scenario $\rho_S \gg p_S$ is symptomatic of cold dark matter, (\ref{2Dflow}) restricts the dynamics of singular shells to surface flows, whereas successful structure formation simulations \citep{Springel2005,Pillepich2018} all rely on dark matter particles moving in a full 3-dimensional space.  More importantly, the spatial shells are spherical concentric, indicative of a preferred position in space, while the {\it same} dark matter, when considered as the cause of decelerated expansion of the entire Universe, is distributed spatially without a center.  

Nevertheless, a different approach from \cite{Lieu2024} which retains the spirit of it has become apparent.  


\section{A \texorpdfstring{$\Lambda$}{Lambda}CDM Universe driven by temporal singularities}

We start with one of the most important tenet of $\Lambda$CDM cosmology, namely local mass-energy conservation in a homogeneous Universe.  In General Relativity it is implied by the vanishing of the covariant derivative of the stress energy tensor, \beq T^{\mu\nu}_{;\nu} =0, \label{T} \eeq where \beq T^{\mu\nu} = (\rho+p) u^\mu u^\nu -pg^{\mu\nu} \label{stress} \eeq  with $g^{\mu\nu}$ being given by (\ref{FRW}), and $\rh$ and $p$ the mass-energy density and pressure of the cosmic subtratum.   Specifically we are referring to the $\mu =0$ case of (\ref{T}), \ie~\beq T^{0\nu}_{;\nu} =0 \implies \dot\rho =-\fr{3\dot a}{a} [\rho (a) + p(a)] \label{continuity} \eeq
which is the mass-energy continuity equation of a fluid undergoing a radial outflow drive solely by cosmic expansion (\ie~no peculiar velocity, $u^i=dx^i/dt=0$).  Hitherto the only known solution to (\ref{continuity}) is that of a single component fluid with $p/\rho = w$, a constant.  It is, assuming the expansion factor $a$ has the value $a_0=1$ at the present time $t_0$, \beq \rho (a)= \fr{\rho_0}{a^{3(1+w)}},~p=\fr{w\rho_0}{a^{3(1+w)}};~\rho_0>0~{\rm and}~a>0, \label{soln1} \eeq and it suits a Universe in which the matter and energy are not only homogeneously distributed, but also omnipresent during times after the initial singularity of the Big Bang, \ie~when $a>0$. 

Although (\ref{soln1}) enforces local mass-energy conservation, it does run into a problem at the cosmic time origin $a=t=0$, because  there has to be an initial creation of particles and quanta at $t=a=0$.  Thus, when (\ref{soln1}) is extended to include the $a=0$ moment, it becomes \beq \rho (a) = \fr{\rho_0}{a^{3(1+w)}} \th(a),~p=\fr{w\rho_0}{a^{3(1+w)}}\th(a);~\rho_0>0~{\rm and}~a\geq 0 \label{soln2} \eeq with $\th(a)$ being the Heaviside step function, which clearly violates (\ref{continuity}) at $t=a=0$.  Now $\Lambda$CDM cosmology sidelines the problem by appealing to quantum fluctuations in spacetime during the initial $\approx 10^{-44}$~s, the Planck time, or the moment of `creation', when there is well established evidence from diffraction limited quasar images of spacetime smoothness on scales beneath the Planck time (\cite{lie03}).  Moreover, all observable quantum effects conserve mass-energy.

However, there is a simple way of restoring consistency between (\ref{soln2}) and (\ref{continuity}), namely by modifying (\ref{soln2}) to include an extra term to the expression for the pressure in (\ref{soln2}), \beq \rho (a) = \fr{\rho_0}{a^{3(1+w)}} \th(a),~p=\fr{w\rho_0}{a^{3(1+w)}}\th(a) -\fr{\rho_0}{3a^{2+3w}} \de(a);~\rho_0>0~{\rm and}~a\geq 0. \label{soln3} \eeq  When (\ref{soln3}) is substituted into the right side of (\ref{FRW}), the ensuing solutions for $\dot a/a$ and $\ddot a/a$ would not differ from (\ref{soln2}), considering the fact the curvature of the very early Universe, \ie~the constant of integration $K$ in (\ref{FRW}), is arbitrary.  This time, in addition to reinstating (\ref{continuity}), (\ref{soln3}) with $1+w\geq 0$ also obeys the weak energy condition, or the condition which ensures positive mass-energy in any measurable volume of space and during any resolvable time interval, namely \beq \<\rho (a)\> \geq 0~{\rm and}~\<\rho (a) +p(a)\> \geq 0 \label{weakcond} \eeq for all $a\geq 0$ and where $\<\cdots\>$ denote averaging over a finite interval of $a$.  The case $a>0$ is obvious; even for $a=0$, during which $\th(a)=1$ but $a\de(a)=0$, one sees that (\ref{weakcond}) still holds because $\rho(a)  >0$ and the ratio $p (a)/\rho (a) \to 0$.  The real difficulty with the Big Bang scenario (\ref{soln3}) is the non-detection of dark matter and dark energy - the indispensable active ingredients which reconcile $\Lambda$CDM cosmology with observations. 

To address the difficulty, it is the purpose of this paper to point out that, in the spirit of (\ref{soln3}), there exists {\it another} solution for $\rho (a)$ and $p(a)$ which also satisfies mass-energy conservation during fluid flow, (\ref{continuity}), and the weak energy condition (\ref{weakcond}).    It is of the form \beq 
\rho (a) = \fr{\xi(\al)}{a^3}  \de(a-\al);~~p(a) =  -\fr{\xi(\al)}{3a^2} \de' (a-\al),~\label{rhoA} 
\eeq 
where $a=a(t)$ is the dimensionless expansion factor and $\xi (\al)$ is an arbitrary function, which is finite and positive for all $\al >0$.  As a result, the density $\rho (a) \geq 0$. 
Evidently (\ref{rhoA}) implies the possibility of $\rho$ and $p$ assuming non-zero values at more than one cosmic time (\ie~at multiple values of $\al$).  Yet because (\ref{rhoA}) obeys (\ref{continuity}) at {\it all} times, it means repeated bursts of matter and energy could occur without the need to create them from nowhere.  
Since the moments of finite $\rho$ and $p$ are short, few, and far between, dark matter and dark energy are {\it not omnipresent}, which explains why it is so hard to find them.

Owing to the behavior of $\delta'(a-\al)$, the pressure to density ratio $w$ fluctuates from $-\infty$ at $a\to \al^+$ to $\infty$ at $a\to \al^-$. 
But there is no violation of causality because even though $|P/\rho| > 1$ the sound speed $c_s = \sqrt{dP/d\rho} \neq \sqrt{P/\rho}$. 
Specifically during the intervals defined by the time shells, matter is released uniformly throughout space, \ie~$\de P=\de\rho=0$ (if perturbations are present to seed structures, the ensuing $\de P$ and $\de\rho$ would be finite, with $|dP/d\rho| < 1$).  Likewise, the space shells of \citep{Lieu2024} are static and do not evolve with time, so that the concern of $c_s > 1$ is as irrelevant as the sound speed itself.

Moreover, during any measurable (\ie~finite, resolvable) time interval containing the instance when $a=\al$, the average of the ratio $3p(a)/\rho (a)$, namely $\<3p(a)/\rho (a)\> = -\<\al\delta'(a-\al)/\de (a-\al)\>$, evidently vanishes because the numerator is an odd function while the denominator is even.  By a similar argument, it is also clear that $\<\rho (a) + p(a)\> \geq 0$.  Since $\<\rho (a) + 3p(a)\> \geq \rho_{\rm min} (a) \< 1+3p(a)/\rho (a)\> = \rho_{\rm min} (a) (1+\<p(a)/\rho (a)\>) = \rho_{\rm min} = 0$, the strong energy condition \beq \<\rho (a)+ p(a)\> \geq 0;~\<\rho (a) + 3p(a)\> \geq 0, \label{strong} \eeq where the brackets $\<\cdots\>$ denote average over a finite interval of $a$ containing $a=\al$, is satisfied.

Concerning the regularization of the delta functions, one does not yet know exactly how singular they are, but the key is that they must be narrow enough to evade direct detection, \ie~in reality one only observes their aggregate effect on space and time as detailed below (width estimates are provided after (\ref{1F}).  This is the only way (\ref{rhoA}) could play a useful role in accounting for the illusiveness of dark matter and dark energy.  
As far as structure formation and stabilization go, they can proceed in the manner of standard simulations, and the flatness-and-horizon problem may still be solved by introducing some form of cosmic inflation.  The point is that neither dark matter nor dark energy has to fill all of space at all times; if they exist only as short intermittent bursts, during which they fill all of space but before and after which they disappear,
they would evade detection but could nevertheless account for the expansion history of the Universe as well as the binding of galaxies, groups, and clusters.  In other words, to account for much of cosmological findings the presence of only temporal shells (\ie~transient particles or quanta) filling all space, is by itself sufficient to offer a unifying account of the key observational phenomena of the Universe, without the need for confinement of the mass-energy by  spatially singular shells (\cite{Lieu2024}) as well. 

One may now construct the Friedmann equations (\ref{Friedmann}) for a homogeneous Universe comprising a set of temporal shells, namely \beq \fr{\ddot a}{a} = -\fr{4\pi G}{3} \sum_j \xi(\al_j) \left[\fr{\de(a-\al_j)}{a^3} -\fr{\de'(a-\al_j)}{a^2}\right], \label{2F} \eeq and 
\beq \left(\fr{\dot a}{a}\right)^2 = \fr{8\pi G}{3} \sum_j \xi(\al_j) \fr{\de(a-\al_j)}{a^3} +\fr{K}{a^2}. \label{1F} \eeq
If the spacing between an adjacent pair of temporal discontinuity in $\rho (a)$ and $p(a)$ is small, $\de a = -\de z/(1+z)^2 \lesssim 0.1$, and current observational data only allow one to measure averaged values of $\ddot a/a$ and $(\dot a/a)^2$, values which match $\Lambda$CDM predictions, \ie~ they will not be able to detect any discreteness in the expansion which may exist on a redshift resolution $\de z \approx \de a \lesssim 0.1$.  The width pf each temporal spike (as given by the delta functions of (\ref{rhoA})) would then be $\de z\approx \de a \lesssim 0.01$, \ie~at least $10$ times narrower than the spacing.

To compute the averaged expansion rates, we start with (\ref{2F}) by converting the right side to an integral over many temporal shells, namely, 
\beq
\begin{split}
\Big\<\fr{\ddot a}{a}\Big\> &= -\fr{4\pi G}{3}\left[\int_0^\infty \fr{\xi(\al)n(\al)}{a^3}\de(a-\al) - \fr{\xi(\al)n(\al)}{a^2}\de'(a-\al)\right]d\al\\
&= -\fr{4\pi G}{3} \left[\fr{\xi(a) n(a)}{a^3}-\fr{1}{a^2}\fr{d}{da} \int_0^\infty \xi(\al)n(\al)\de(a-\al) d\al\right]\\
&= -\fr{4\pi G}{3} \left\{\fr{\xi(a) n(a)}{a^3}-\fr{1}{a^2}\fr{d}{da} [\xi(a)n(a)]\right\}. \label{int2F} 
\end{split}
\eeq
where $n(\al) d\al$ is the number of temporal shells between expansion factors $\al$ and $\al+d\al$, and the limits of integration should be $\al_{\rm min}$ and $\al_{\rm max}$ if the episode of space expansion driven by the set of shells under consideration lasts only a finite interval of time (assuming, as usual, that $a=a(t)$ where $t$ is the cosmic time).  Similarly, one can integrate (\ref{1F}) to obtain \beq \Big\< \left(\fr{\dot a}{a}\right)^2 \Big\rangle = \fr{8\pi G}{3} \fr{\xi(a)n(a)}{a^3} + \fr{K}{a^2}. 
 \label{int1F} \eeq 
How the Hubble parameter $(\dot a/a)^2$ and the acceleration parameter $\ddot a/a$ evolves with cosmic time $t$ would depend on the exact functional form of $\xi (a)$ and $n(a)$.

By comparing (\ref{2F}) and (\ref{1F}) against (\ref{Friedmann}), one infer the equivalent (or effective) density and pressure of the system of temporal shells, as \beq \rho_S (a) = \fr{\xi(a) n(a)}{a^3};~p_S (a) =  -\fr{1}{3a^2}\fr{d}{da} [\xi(a)n(a)]. \label{denpre2} \eeq 
Note that $\rho_S (a)$  and $p_S (a)$ also obey the full 3-dimensional continuity equation 
in the form given by (\ref{continuity}), because the shells here are in the {\it time} dimension, \ie~there are no surface structures expanding in space, only entire volumes expanding homogeneously in three dimensions.  As to the constraints on $\xi (a) n(a)$ to ensure a stable thermal history of the Universe, it will be shown below, through the argument given below (\ref{crit}), that they are, for power law evolution, $\xi(a) >0$ and $\xi (a) n(a) \propto a^{3\ep}$ where~$\ep \geq 0$.  


We now demonstrate self-consistency in the above expressions for $(\dot a/a)^2$ and $\ddot a/a$.  Starting with the latter, we may integrate (\ref{2F}) once w.r.t. $a$ to obtain 
\bea
    \tfrac{1}{2} \dot a^2 & =&  \int \ddot a da \\
&=&  -\fr{4\pi G}{3} \sum_j \xi(\al_j) \int\left[\fr{\de (a-\al_j)}{a^2} - \fr{\de'(a - \al_j)}{a}\right]da +\fr{K}{2} \\
&=& \fr{4\pi G}{3} \sum_j \fr{\xi(\al_j)}{a} \de (a-\al_j) +\fr{K}{2}, \label{H}
\eea
where $K$ is a constant of integration.  Next, assuming that there are many successive temporal shells, one may convert the summation over $j$ to an integral over $\al$ as before, resulting in (\ref{1F}).  

From (\ref{int2F}), (\ref{int1F}), and (\ref{denpre2}) it is evident that an expansion of space driven predominantly by dark matter and dark energy would correspond to the scenario \beq \xi_m (a) n_m (a) = \rho_0,~{\rm or}~\xi_\Lambda (a) n_\Lambda (a)= \rho_\Lambda a^3 \label{DMDE} \eeq where $\rho_0$ is a constant, being the density of dark matter at redshift $z=0$ when the expansion factor is $a(t_0) =a_0 =1$, and $\rho_\Lambda$ is also a constant, being the density of dark energy.  The total matter density at $z=0$ is then $\rho_0 + \rho_b$ while the density of matter and dark energy combined is $\rho_0 + \rho_b + \rho_\Lambda$ at $z=0$, where $\rho_b$ is the present
baryon mass density (including radiation).

\section{Flatness, structure formation, inflation}

In the previous section it was demonstrated that the physical consequence of a time sequence of successive and sudden release and disappearance of particles across all space in a manner consistent with mass-energy conservation could, if temporally unresolved by an observer, manifest itself as a continuous expansion of space with properties symptomatic of dark matter, dark energy, or matter having some other equation of state.  Moreover, (\ref{2F}) and (\ref{1F}) indicate that the total amount of mass-energy in these transient events, when integrated over a range of expansion factor $a(t)$, is the same as the scenario of genuine continuous expansion of space, after it is initially and homogeneously filled with matter. 

Thus, the bulk of $\Lambda$CDM cosmology remains robust in this model.  Structure formation may be pursued by spatially perturbing some cosmological parameters, exactly which ones would depend on the 
choice of gauge \citep{Peebles1980,Mukhanov1992}.  Flatness and horizon problems, the former implying a small $K$ in (\ref{1F}),  may still be solved by postulating a period of exponential expansion, \ie~inflation, in the early Universe, except that the current approach allows inflation to commence or complete over any finite time interval without the need to dissipate away any scalar field.  This is because each individual temporal shell (\ref{rhoA}) separately satisfies the continuity equation, \ie~there could conceivably be a group of shells satisfying the second of (\ref{DMDE}), and spanning some finite range of $a$ to mark the start and end times of inflation without producing any other physical problems, such as how the inflaton scalar field was eventually dissipated to replenish the original content of the Universe.  

In view of this realization, and the understanding that superhorizon perturbations do not grow in the Newtonian (or longitudinal) gauge unless it enters the horizon at matter-radiation equality or later \citep{Mukhanov1992}, it seems that a simpler approach to inflation is to assume that it occurred at around equality ($z_{\rm eq} \approx$~3,400,~age $t_{\rm eq} \approx 1.5\times 10^{12}$~s) and lasted at least $\ln (t_0/t_{\rm eq}) \approx 12.5$ e-folds, or a duration $12.5 t_{\rm eq} \approx 2\times 10^{13}$~s, such that the horizon scale at equality is inflated to match the present horizon or beyond, \ie~such that $t_{\rm eq} e^{t/t_{\rm eq}} = t_0$ where $t_0\approx 4.2\times 10^{17}$~s is the present age of the Universe.  The temperature of the Universe during inflation would drop by the same 12.5 e-folds, as $T \propto a^{-1}$.  Since the expected temperature of the Universe at equality is $\approx 10^4$~K, that means the temperature before inflation is about 2.8~$\times 10^9$~K.  Moreover, a 12.5 e-fold or longer inflation period implies that a radius of curvature at $t=t_{\rm eq}$ of order the Hubble radius $1/H_{\rm eq}$ would, after inflation, become $\approx 2.8\times 10^5$ smaller than the present Hubble radius, which is well within the observational upper limit for the curvature parameter $K$.  This solves the horizon problem very simply, without invoking any inflaton dissipation mechanisms \citep{Kofman1994,Kofman1997,Brandenberger1999}.  

The merit in pushing the inflationary epoch to a much later time than the conventional $t\approx 10^{-35}$~s is that, if the post-inflationary thermal Universe in the absence of dissipative reheating is one which would evolve to $z=0$ having current observational properties, the temperature of the Universe at the Planck time $t\approx 5\times 10^{-44}$~s would have to exceed the Planck temperature $1.42\times 10^{32}$~K unless inflation took place far later than $10^{-35}$~s.


To be quantitative about the seeding of structure formation, one returns to the perturbed zero curvature\footnote{It is assumed that any initial curvature is removed by an episode of inflation (see the last paragraph).} FRW line element, of the general form \beq ds^2= (1-2\Psi) dt^2 -2aB_{,i} dt dx^i - a^2 [(1-2\Phi) \de_{ij} +2E_{,ij}] dx^i dx^j. \label{pertds} \eeq Ignoring shear and anisotropic effects, one would then ignore the $B_{,i}$ and $E_{,ij}$ terms and set $\Psi=\Phi$.  In this case, it was shown in previous literature (see \eg~the two references cited below) that the total curvature perturbation of the inflaton and radiation energy density, expressed in Fourier ($k$) space and denoted by $\rho_I$ and $\rho_r$ respectively, such that the stress-energy tensor of the Universe in the inflationary era is given by (\ref{stress}) with $\rho=\rho_I + \rho_r$ and $g^{\mu\nu}$ as in (\ref{pertds}), namely 
\beq 
\zeta = -\Phi - H\fr{\de\rho}{\dot\rho}, \label{zeta}
\eeq
is gauge invariant even on superhorizon scales.   In (\ref{zeta}), $H=\dot a/a$, and $\Phi$ is related to the density perturbation (or perturbation in $\xi (a)$ in the second of (\ref{DMDE})) by \beq \dot\Phi + H\Phi = -\fr{H}{2}\fr{\de\rho}{\rho}. \label{Phi} \eeq Now, in the Newtonian gauge $\zeta$ does not change with time  while a perturbation mode is outside the horizon, \ie~while $k < 1/H$.  It was shown (see \citep{mal03,lie23}) that, in respect of standard inflation, if $\zeta$ assumes the same constant value independent of $k$ for all $k< 1/H$, \ie~if the small $k$ power spectrum of $\zeta$ is flat in the Newtonian gauge, all superhorizon modes will re-enter the horizon after post-inflationary reheating (when the thermal Universe is reborn) with the same perturbation amplitude, albeit at different times with the modes of smaller $k$ re-entering later.

All this applies to conventional theory.  The question we must address is whether it also works when the inflaton field as given by something close to the second of (\ref{DMDE}), which can disappear after the last time shell comes to pass, or gradually fade away in strength from one shell to the next without dissipating into radiation.  More precisely, if one defines the parameter $\ep=1+w_I$, where $w_I=p_I/\rho_I$ is the equation of state index of the inflaton: $\ep=0$ would reduce the inflaton field to dark energy, $0<\ep\ll 1$ marks the the era of inflation, while $\ep > 4/3$ would mark the gradual end of inflation unless there is a sudden disappearance of the corresponding time shells.  Provided \beq \rho_I \geq 0~{\rm and}~\ep\geq 0 \label{crit} \eeq  which is {\it always} the case in the time shell model, the weak energy condition is universally satisfied, see \citep{shl24}.  The quantity $\xi (a) n_I (a)$ of (\ref{DMDE}) would then trend with $a$ as $\xi(a) n(a) \propto a^{3\ep}$, or \beq \xi_I (a) n_I (a) = \rho_{I0} a^{3\ep}, \label{inflaton} \eeq where the subscript I stands for the inflaton.  

Now it was shown in\citep{mal03,lie23} that the evolution of $\zeta$ during inflation is given by a pair of equations, namely  \beq \dot\zeta = \fr{3H}{\dot\rho}[(w_I-w_r)\dot\rho_I + \dot\ep\rho_I] (\zeta -\zeta_I) \label{dotzeta1}, \eeq or \beq \dot\zeta = \fr{3H}{\dot\rho}(w_r-w_I)\dot\rho_r (\zeta -\zeta_r) \label{dotzeta2}, \eeq where $w_r = 1/3$ for radiation, either of which is a valid equation for $\dot\zeta$.  After some e-folds of inflation, however, $\rho_I \gg\rho_r$ and so $\zeta\approx\zeta_I$; the first relation ensures $\dot\zeta\approx 0$ by (\ref{dotzeta2}) while the second ensures $\dot\zeta\approx 0$ by (\ref{dotzeta1}).  If inflation ends with neither inflaton dissipation nor the violation of energy conservation (as explained earlier, only the time shell solution of the mass-energy continuity equation can achieve this), which can take place by a rapid increase in $\ep$ or by the notion of a `last time shell', then, at all subsequent times when the cosmic substratum comprises radiation (with $\rho_r \gg \rho_b \gg \rho_I$), then, (\ref{dotzeta1}) becomes irrelevant but (\ref{dotzeta2}) still yields a vanishing $\dot\zeta$ because $\zeta$ is now given by $\zeta =\zeta_r$.  The conclusion is therefore that the same curvature perturbation $\zeta$ which existed during inflation is also found in the post-inflationary thermal Universe, without any dissipation mechanism necessary.

\section{Conclusion}

Motivated by an earlier work on concentric spherical shells of density singularity, which give rise to zero integrated total mass but a $1/r$ on-shell attractive force \citep{Lieu2024}, this paper proposes an improved version of the model, which is also radically different. The new model can account for both structure formation and stability, {\it and} the key observational properties of the expansion of the Universe at large, by enlisting density singularities in {\it time} that uniformly affect all space to replace conventional dark matter and dark energy. 
Whether the temporal singularities mimic dark matter or dark energy when their effects are manifested in continuum form to an observer unable to resolve them would depend on the interplay between the number density of the singularities and their amplitude.  The space shells of \citep{Lieu2024} may still be responsible partly for the binding of structures, and fully for the existence of giant rings and walls on Gpc scales cited in \citep{Lieu2024}.


Another notable advantage of the model is that because each individual temporal singularity satisfies the continuity equation, it can appear and disappear at arbitrary discrete cosmic times, with arbitrary amplitude.  For singularities which mimic dark energy, this means they can be present within a certain finite interval of cosmic history, \ie~one does not have to account for the `rise and fall' of dark energy by (as is usually done) postulating a prehistoric scalar field which eventually dissipates into thermal particles (\citep{Kofman1994,Kofman1997,Brandenberger1999}).   The consequence for inflationary cosmology is that the flatness and horizon problem, as well as primordial density perturbations which seeded density perturbations, could all be solved rather simply by postulating 15 e-folds of exponential expansion as driven by a group temporal singularities located at about the time of matter-radiation equality.  This means that before inflation the Universe was radiation dominated by $12-13$ e-folds.

The physical origin of the proposed temporal singularities is currently unknown, as topological defects of the early Universe (\citep{Kibble1976}) only give rise to domains, walls, and cosmic strings, which are discontinuities in space but not in time.  Nonetheless, the singularities are at least viable in the sense that they satisfy mass-energy conservation and the strong energy condition.  Moreover, the ideas presented in this paper are not any more presumptuous
standard $\Lambda$CDM, which predicates on the temporal singularity of the Big Bang as the moment of creation, so one must also acknowledge that the ultimate physical origin of $\Lambda$CDM cosmology is, ahead of the availability of a robust quantum gravity theory, likewise unknown.  The only difference between this work and the standard model is that the temporal singularity occurred only once in the latter, but more than once in the former. 
\label{lastpage}

\end{document}